\documentclass[conference]{IEEEtran}

\IEEEoverridecommandlockouts
\usepackage{cite}
\usepackage{amsmath,amssymb,amsfonts}
\usepackage{algorithmic}
\usepackage{graphicx}
\usepackage{subcaption}
\usepackage{textcomp}
\usepackage{subcaption} 
\usepackage{xcolor}
\usepackage{soul}
\usepackage{float}

\usepackage[hyphens]{url}
\usepackage{hyperref}
\hypersetup{breaklinks=true}

\def\AA{\mathcal{A}}
\def\PP{\textbf{P}}

\usepackage{amsthm} % proofs
\usepackage[final]{changes}

\newtheorem{condition}{Condition}
\newtheorem{criterion}{Criterion}

\makeatletter
\newcommand{\linebreakand}{%
  \end{@IEEEauthorhalign}
  \hfill\mbox{}\par
  \mbox{}\hfill\begin{@IEEEauthorhalign}
}

\begin{document}

\title{Unpacking Maximum Extractable Value on Polygon: A Study on Atomic Arbitrage
% Maximum Extractable Value (MEV) in Polygon Network
% \thanks{Identify applicable funding agency here. If none, delete this.}
}

\author{\IEEEauthorblockN{
Daniil Vostrikov\IEEEauthorrefmark{1},
Yash Madhwal\IEEEauthorrefmark{2},
Andrey Seoev\IEEEauthorrefmark{3},
Anastasiia Smirnova\IEEEauthorrefmark{3}, 
Yury Yanovich\IEEEauthorrefmark{2},
Alexey Smirnov\IEEEauthorrefmark{3}, \\
Vladimir Gorgadze\IEEEauthorrefmark{1}
}
\IEEEauthorblockA{\IEEEauthorrefmark{1}%
  Moscow Institute of Physics and Technology, Moscow, Russia \\
}
\IEEEauthorblockA{\IEEEauthorrefmark{2}%
  Skolkovo Institute of Science and Technology, Moscow, Russia \\
}
\IEEEauthorblockA{\IEEEauthorrefmark{3}%
  MEV-X, Moscow, Russia 
}
}

\maketitle

\begin{abstract}
% v2
The evolution of blockchain technology, from its origins as a decentralized ledger for cryptocurrencies to its broader applications in areas like decentralized finance (DeFi), has significantly transformed financial ecosystems while introducing new challenges such as Maximum Extractable Value (MEV). This paper explores MEV on the Polygon blockchain, with a particular focus on Atomic Arbitrage (AA) transactions. We establish criteria for identifying AA transactions and analyze key factors such as searcher behavior, bidding dynamics, and token usage. Utilizing a dataset spanning 22 months and covering 23 million blocks, we examine MEV dynamics with a focus on Spam-based and Auction-based backrunning strategies. Our findings reveal that while Spam-based transactions are more prevalent, Auction-based transactions demonstrate greater profitability. Through detailed examples and analysis, we investigate the interactions between network architecture, transaction sequencing, and MEV extraction, offering comprehensive insights into the evolution and challenges of MEV in decentralized ecosystems. These results emphasize the need for robust transaction ordering mechanisms and highlight the implications of emerging MEV strategies for blockchain networks.
\end{abstract}


\section{Introduction}\label{sec:Intro}

% Remove it later
% \noindent\rule[0.5ex]{\linewidth}{1pt}
% \textbf{!!NOTE!!}
% \todo{\textbf{Red} text to be rewritten.}
% \notes{\textbf{Blue} text to be \st{deleted}.}
% \noindent\rule[0.5ex]{\linewidth}{1pt}

The rapid adoption of blockchain technology has revolutionized financial ecosystems, enabling innovations such as decentralized finance (DeFi) protocols and non-fungible tokens (NFTs) \cite{BanaeianFar2023,razi2023non,boreiko2024decentralized}. However, its transparency also introduces Maximal Extractable Value (MEV), which refers to the value block producers (miners or validators) can capture by reordering, including, or excluding transactions within a block \cite{Gramlich2024, Materwala}. MEV arises from the openness of blockchain systems, where pending transactions are visible in the mempool, allowing block producers to manipulate transaction order for profit.

MEV presents both technical and ethical challenges, influencing gas markets, transaction fairness, and blockchain security \cite{sarkar2023fairflow, alipanahloo2024maximum, mazorra2022price}. Miners and validators, often collaborating with arbitrage bots, extract MEV. Platforms like Flashbots aim to mitigate the negative effects of MEV but have raised concerns about centralization and transparency \cite{lysaght2022exploring,ji2024regulatory}. Understanding MEV is crucial for balancing economic incentives with decentralization and fairness.

Initially observed on Ethereum, MEV began with simple arbitrage across decentralized exchanges (DEXs) but later evolved into more complex strategies such as frontrunning, backrunning, and sandwich attacks \cite{zecirovic2024analysis,zust2021analyzing,shou2024backrunner}. By 2020, the rise of searchers and bots further intensified MEV extraction, prompting platforms like Flashbots to reduce gas wars and improve transparency \cite{Flashbot}. Despite these efforts, researchers warn that MEV may destabilize blockchain consensus and drive innovation in slippage protection, fair ordering protocols, and MEV-aware consensus algorithms \cite{li2023demystifying,chaurasia2024mev,barczentewicz2023mev,mancino2023exploiting}.

While most MEV research has focused on Ethereum, there remain gaps in understanding how other blockchain networks, such as Polygon, handle transaction ordering and economic incentives. This paper examines MEV on Polygon, focusing on Atomic Arbitrage (AA) transactions across Uniswap V2, Uniswap V3, Algebra, and partially covering Balancer \cite{uniswapv2,uniswapv3,algebra,balancer}. We analyze Spam-based and FastLane-based backrunning strategies over a 22-month period (January 2023 to October 2024) \cite{fastlane}, providing insights into MEV extraction dynamics.

The main contributions of this paper are a detailed case study of AA transactions on the Polygon blockchain, the introduction of criteria for identifying and analyzing AA, and a numerical analysis of MEV focused on Spam-based and FastLane-based AA. These findings highlight distinct operational characteristics and profitability trends, offering insights to optimize blockchain protocols for fairness, efficiency, and decentralization.

The paper is structured as follows. Section~\ref{sec:background} covers MEV taxonomy and arbitrage mechanisms. Section \ref{sec:RelatedWork} reviews prior studies and research gaps. Section \ref{sec:PolygonArb} provides Polygon-based MEV examples, while Section \ref{sec:heuristic} presents AA transaction identification criteria. Section \ref{sec:Experiments} outlines data collection methods, and Section \ref{sec:RnD} discusses results and their implications. Finally, Section~\ref{sec:Conclusion} concludes and proposes future research directions.

\section{Background}\label{sec:background}

\subsection{MEV Taxonomy}
MEV is the additional value validators or miners can capture by reordering, including, or censoring transactions in a blockchain block, particularly in networks like Ethereum \cite{tjiam2021your}. Due to the transparency of pending transactions, validators can manipulate transaction sequences for profit. MEV can be categorized as follows:

\begin{enumerate}
    \item \textbf{Atomic MEV:} Captured within a single transaction, ensuring guaranteed profit through:

    \begin{itemize}
    
        \item \textbf{Arbitrage:} Exploiting price discrepancies across DEXs.

        \item \textbf{Liquidations:} Acquiring discounted collateral from undercollateralized positions.

        \item \textbf{Long-Tail MEV:} Less frequent strategies exploiting niche inefficiencies.
    \end{itemize}
\end{enumerate}

\begin{enumerate}
    \item \textbf{Semi-Atomic MEV:} Requires precise timing but may span multiple transactions.
    \begin{itemize}
        \item \textbf{Sandwich Attacks:} Placing trades around a target transaction to capitalize on price movement.
        \item \textbf{Multi-Layer Sandwich Attacks:} \added{An advanced form of sandwiching where multiple intermediate swaps or liquidity layers are used across different protocols to extract value, typically involving complex coordination and route fragmentation.}
        \item \textbf{Sandwich Arbitrage:} Combining sandwich attacks with arbitrage for increased profit.
    \end{itemize}

    \item \textbf{Statistical Arbitrage:} Uses predictive models to exploit price inefficiencies. Profiting from price differences between centralized exchanges (CEXs) and DEXs.

    \item \textbf{Intent-Based Swaps (IBS):} Maximizing profits through optimal ordering of solver transactions.
    \begin{itemize}
        \item \textbf{OppTx + FastLane Atlas TX:} Bundling transactions for efficient execution.

        \item \textbf{IBS Protocols:} Platforms like CoW Swap, 1inch Fusion, and UniswapX facilitate such strategies.
    \end{itemize}
\end{enumerate}

\subsection{Arbitrage and Opportunity Analysis}
Arbitrage exploits price variations across liquidity pools, enabled by blockchain transparency. MEV opportunities arise due to:

\begin{itemize}
    \item \textbf{Market Dynamics:} Rapid price changes and liquidity variations.
    \item \textbf{Transaction Latency:} Delays in price updates between exchanges.
    \item \textbf{Mempool Transparency:} Visible pending transactions allow frontrunning and backrunning.
    \item \textbf{Protocol-Specific Mechanics:} Differences in fee structures and slippage tolerances.
\end{itemize}

Backrunning plays a key role by capturing residual value from previous transactions and improving market efficiency.

\begin{figure}
    \centering
    \includegraphics[width=\linewidth]{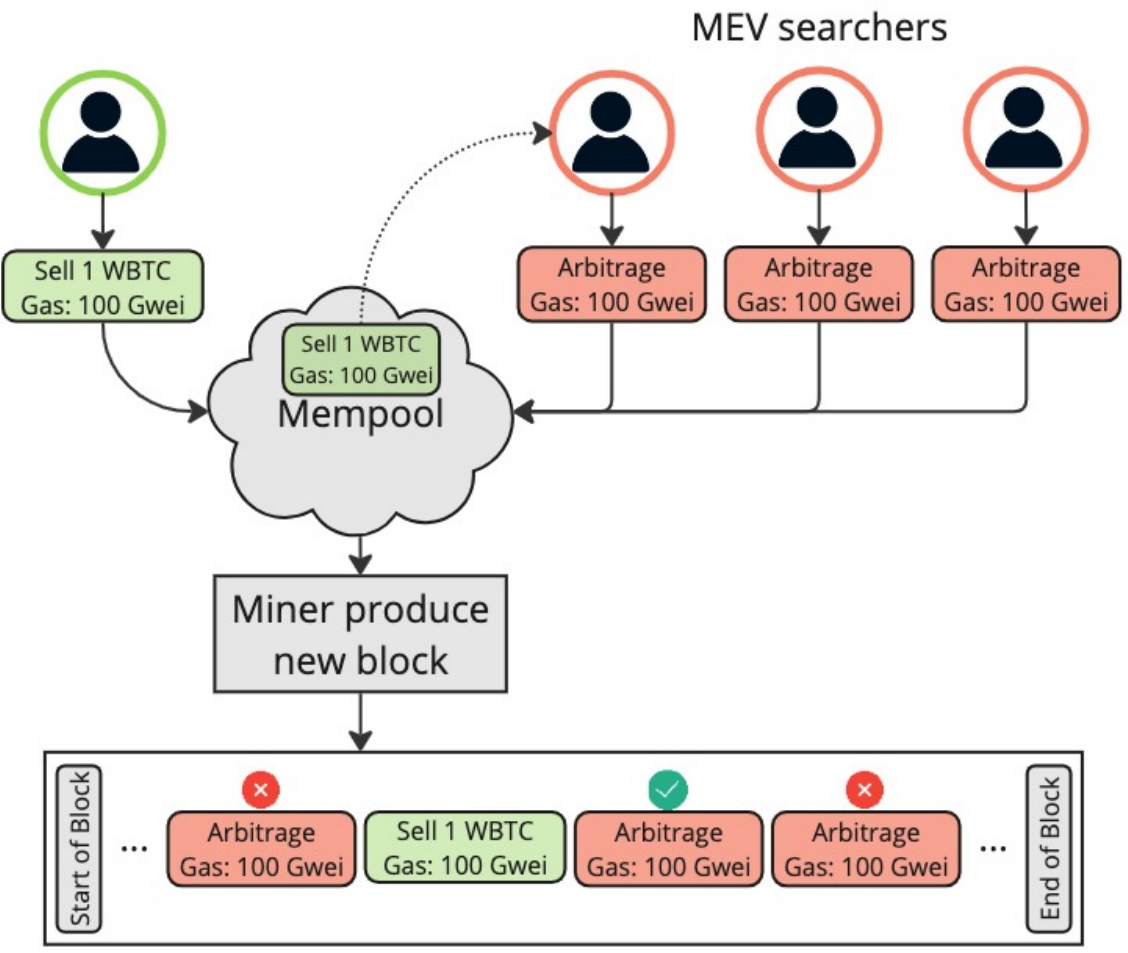}
    \caption{Successful backrun transaction placed immediately after the opportunity transaction to capture MEV opportunity to buy 1WBTC.}
    \label{fig:backrun}
\end{figure}

\subsection{Backrunning}
Backrunning involves placing a transaction immediately after a target transaction to capture value (see Figure \ref{fig:backrun}). Two main approaches exist:

\textbf{Spam-Based:} Broadcasting multiple identical transactions to increase chances of placement, often causing network congestion and high gas fees \cite{Kruglik2019}.

\textbf{Auction-Based:} A structured system using private transaction relays and bidding auctions to minimize inefficiencies.
% FastLane optimizes MEV extraction while reducing network congestion.

\subsection{Atomic Arbitrage (AA)}
AA executes interdependent swaps in a single transaction, eliminating partial execution risk. The strategy ensures:

\begin{itemize}
    \item All swaps occur within a single atomic transaction.
    \item The transaction results in net profit after fees.
    \item Cross-market interactions exploit price discrepancies.
    \item Failure in any step reverts the entire transaction.
\end{itemize}

This approach is crucial in DeFi, minimizing risk while maximizing efficiency across liquidity pools.

\section{Related Work}\label{sec:RelatedWork}

\added{The first documented study of MEV on the Polygon network was introduced by the Flashbots team in 2021~\cite{flashbots2021polygon}, offering early insights into MEV extraction behaviors, including arbitrage patterns and the absence of private relay infrastructure. While preliminary, it laid the groundwork a for subsequent in-depth analysis.} Extensive research has examined MEV phenomena such as frontrunning, sandwich attacks, and arbitrage within Ethereum’s ecosystem, primarily using tools like Flashbots. However, studies on MEV dynamics in other blockchains remain limited.

Weintraub et al. \cite{Weintraub2022} analyze MEV’s impact on network fairness, decentralization, and transaction costs, highlighting the role of Flashbots in mitigating gas price wars while raising centralization concerns. Despite democratizing MEV access, Flashbots reinforce the dominance of large mining pools, necessitating more decentralized solutions. Torres et al. \cite{Torres2021FrontrunnerJA} classify frontrunning attacks on Ethereum, identifying displacement, insertion, and suppression as key strategies. Their five-year study, covering 11 million blocks, reveals nearly 200,000 attacks with \$18.41 million in profits, underscoring the need for improved transaction ordering. Tumas et al. \cite{Tumas2023} explore frontrunning in the XRP Ledger, demonstrating that its pseudo-random transaction sequencing remains vulnerable to sophisticated attacks. Their findings emphasize the necessity for better countermeasures to protect users from manipulation. Bagourd et al. \cite{Bagourd} investigate MEV in Layer 2 (L2) networks, particularly on Polygon, Arbitrum, and Optimism. They identify AA as the dominant MEV type and analyze Polygon’s reliance on Priority Gas Auctions (PGAs), which foster spamming. At the same time, Arbitrum and Optimism’s single-sequencer models shift MEV strategies toward latency-driven backrunning. \added{Yang et al. \cite{Yang2024} provide a comprehensive systematization of MEV countermeasures, categorizing 32 proposals into four classes: auction-based solutions (e.g., MEV-Boost, Flashbots), time-based fairness (e.g., Aequitas, Themis), content-agnostic ordering (e.g., TEX, Shutter Network), and application-layer mitigations (e.g., CoW Swap, FairTraDEX). While auction-based approaches such as FastLane partially align with these strategies, Polygon’s architecture lacks native support for time-based ordering or threshold encryption, limiting the deployment of more robust consensus-layer defenses. However, application-level solutions like CoW Swap and intent-based DEXs may offer deployable mitigations on Polygon without modifying the protocol layer. As such, Polygon presents a practical setting to explore how decentralized applications can internalize MEV defenses without enshrined ordering guarantees.} \added{Recent work by Bahrani et al. \cite{SAKA} introduces the SAKA mechanism, a strategy-proof auction-based transaction sequencing protocol that aligns incentives for users, searchers, and block producers. Unlike Flashbots or FastLane, SAKA ensures truthful bidding and fair ordering through a protocol-native design. While not yet implemented on sidechains like Polygon, it demonstrates that incentive-compatible auction mechanisms are both feasible and actively evolving.} Building on this research, our study focuses on AA in Polygon, particularly the impact of FastLane updates, an aspect unexplored in previous work. Analyzing these evolving strategies, we provide insights into how AA adapts to protocol changes, shaping MEV extraction in decentralized finance.

\section{Examples of Atomic Arbitrage on Polygon}\label{sec:PolygonArb}

This section offers examples of notable AA transactions, providing real-world illustrations to support the criterion presented in the subsequent section.

\subsection{Plain Arbitrage}

\begin{figure}[h]
    \centering
    \includegraphics[width=0.9\linewidth]{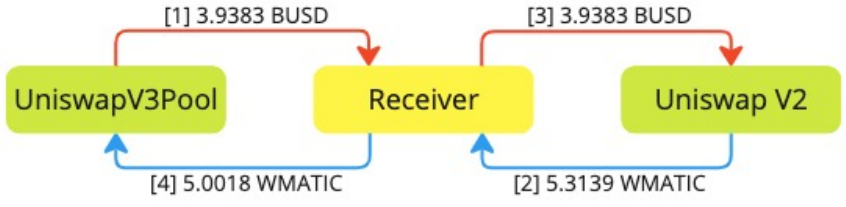}
    \caption{Plain arbitrage between Uniswap V3 and V2 on Polygon netted 0.3121 WMATIC profit (transaction \texttt{0xc6591b9..} in block 58,329,504)}
    \label{fig:plain}
\end{figure}

The Figure \ref{fig:plain} is the transaction that exemplifies a straightforward arbitrage operation on the Polygon network, where the trader exchanged 3.9383 BUSD for 5.3139 WMATIC \cite{Polygonscan2024}. The operation utilized both Uniswap V3 and Uniswap V2 liquidity pools, with Uniswap V3 being used to exchange BUSD for WMATIC by leveraging its concentrated liquidity for efficient price execution. Conversely, Uniswap V2 facilitated the reverse exchange, where WMATIC was swapped back for BUSD. The arbitrage profit was realized when the trader sent only 5.0018 WMATIC back to Uniswap V3 while retaining the difference of approximately 0.3121 WMATIC as profit. By identifying that WMATIC was undervalued on the DEX relative to its actual market value, the trader capitalized on this price discrepancy by purchasing WMATIC at a lower price on one platform and selling it at a higher price on another, thus securing a profit. This type of arbitrage leverages inefficiencies in market pricing and highlights the dynamic interplay of liquidity and valuation across blockchain ecosystems, ensuring better price equilibrium while enabling traders to extract value effectively.

\subsection{120 000\$ trade volume Transaction}

\begin{figure}[h]
    \centering
    \includegraphics[width=\linewidth]{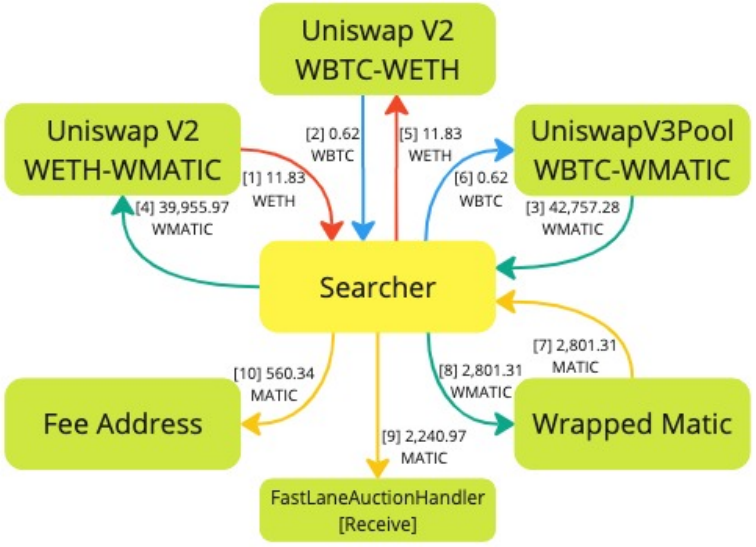}
    \caption{Transaction achieving a trading volume of 120,000 USD (transaction \texttt{0xd331d7..} in block 55,181,032)
     }
    \label{fig:tradeVolume}
\end{figure}

The Figure \ref{fig:tradeVolume} illustrates an arbitrage transaction in the Polygon network, demonstrating the efficiency and profitability of leveraging liquidity pools across platforms. This transaction achieved an impressive trading volume of \$120,000 in a single operation and generated a profit of 2,801 WMATIC \cite{Polygonscan2024}. Utilizing a FastLane mechanism with a length of 3, the trader swapped 39,955 WMATIC into 42,757 WMATIC, resulting in a net profit of 2,801 WMATIC. The operation involved multiple liquidity pools and assets. Initially, the trader interacted with QuickSwap’s WMATIC-WETH pool, exchanging 42,757.3 WMATIC for 11.8 WETH. Concurrently, they received 0.62 WBTC and 42,757.3 WMATIC in their wallet from different liquidity providers. Subsequently, the trader returned 39,956 WMATIC to QuickSwap’s WMATIC-WETH pool while retaining the surplus as profit. The transaction loop was completed with the trader returning 11.8 WETH and 0.62 WBTC to the respective liquidity providers. The profit margin was facilitated by the FastLane mechanism, which efficiently managed the swaps and minimized slippage. By interacting with the smart contract, the trader ensured the optimal execution of the arbitrage strategy. This transaction highlights the advanced capabilities of automated arbitrage systems in blockchain ecosystems, exploiting price discrepancies and dynamic liquidity to achieve high profitability and efficient capital utilization.

\section{Criterion for Identifying Arbitrage Transactions}\label{sec:heuristic}

Arbitrage transactions in DeFi involve executing multiple swaps to exploit price differences across various liquidity pools or assets. To systematically identify these opportunities, we propose a heuristic that evaluates transaction inputs, outputs, and profitability.

\begin{condition}[Multi-Swap]
    A transaction must include at least two swaps:
    $N \geq 2,$
    where $N$ is the number of swaps in the transaction.
\end{condition}

Condition 1 ensures that arbitrage requires sequential trades between different assets or pools to capture price inefficiencies.

Consider $\AA= \{X, Y, Z, \dots\}$ as the set of assets involved in the transaction. The net change for an asset $A \in \AA$ is calculated as:
\[ \Delta(A) = \sum_{n=1}^{N} {\delta_{n,A}},\]
where $\delta_{n,A}$ represents the change in asset $A$ during swap $n$.

\begin{condition}[Sufficiency]
    The net change for every asset $A \in \AA$ involved in the transaction must be non-negative after all swaps: $\Delta(A) \ge 0$.
\end{condition}

Condition 2 ensures that the transaction does not result in a loss of value for any asset.

Transaction profitability is defined as:
\[ \texttt{Profit} = \sum_{A \in \AA} \Delta(A) \cdot \PP(A) - \tau - \beta,\]
where:
\begin{itemize}
    \item $\PP(A)$ is the price of asset $A$.
    \item $\tau$ represents transaction fees.
    \item $\beta$ accounts for bids paid for prioritization.
    \item All prices are expressed in a common currency, such as MATIC.
\end{itemize}

A transaction is considered profitable if $\texttt{Profit} > 0$.

\begin{condition}[Profitability]
For a transaction to qualify as arbitrage, it must yield a positive net profit after accounting for all costs, including transaction fees and prioritization bids: $\texttt{Profit} > 0$.
\end{condition}

We introduce our heuristic criterion for identifying AA:
\begin{criterion}[Atomic Arbitrage]\label{caa}
    A transaction is classified as AA if and only if it satisfies Conditions 1-3, meaning it is a sufficiently profitable multi-swap.
\end{criterion}

\section{Data collection}\label{sec:Experiments}

This section examines MEV transactions in the Polygon network using data collected from a remotely hosted Polygon node. The dataset comprises approximately 23 million blocks spanning 22 months, from January 2023 to October 2024. The MEV volume is analyzed weekly to identify trends and variations over time, with the results presented in Section \ref{sec:RnD}. The experiment is structured in two phases: data collection and subsequent analysis.

The data for this study was collected from a remotely hosted Polygon archive node, accessed via an SSH API. In addition, the Polygon Scanner API was utilized to supplement and verify the data. Python scripts were employed, leveraging the web3py library to parse transactions from the dataset spanning the period by traversing blocks sequentially, starting from October 2024 and moving backward to January 2023. 

In the analysis of AA transactions in the Polygon network, we identified certain transactions that, while appearing to be legitimate arbitrage, were actually the result of a hack on the KyberSwap platform \cite{SlowMist2023,Wang2024a}. Specifically, five transactions that occurred on November 23, 2023, were removed from our dataset. These transactions were linked to the exploit where an attacker stole approximately 54.7 million USD due to a vulnerability in the Reinvestment Curve feature of KyberSwap's Elastic pool. By excluding these transactions, we ensure that our analysis accurately reflects genuine arbitrage activities and not those arising from malicious exploits.

% v2
Following data collection, AA transactions were identified according to Criterion~\ref{caa}. Additionally, all FastLane transactions were gathered. The intersection of FastLane and AA transactions is termed as FastLane-based. The remaining AA transactions are classified as Spam-based, while the remaining FastLane transactions refer to MEV activities that are beyond the scope of this paper. Although Spam non-AA MEV transactions exist, they are not included in our data.

\section{Results and Analysis}\label{sec:RnD}

The following analysis presents key trends in AA activity, focusing on transaction volumes, MEV extraction, searcher participation, and bid efficiency. All figures depict data aggregated over consecutive 4-week intervals, providing a structured view of evolving dynamics within the Polygon ecosystem. This timeframe allows us to capture meaningful shifts in market behavior while maintaining consistency in trend evaluation.

\subsection{Numerical Results}

\begin{figure}[h]
    \centering
    \includegraphics[width=\columnwidth]{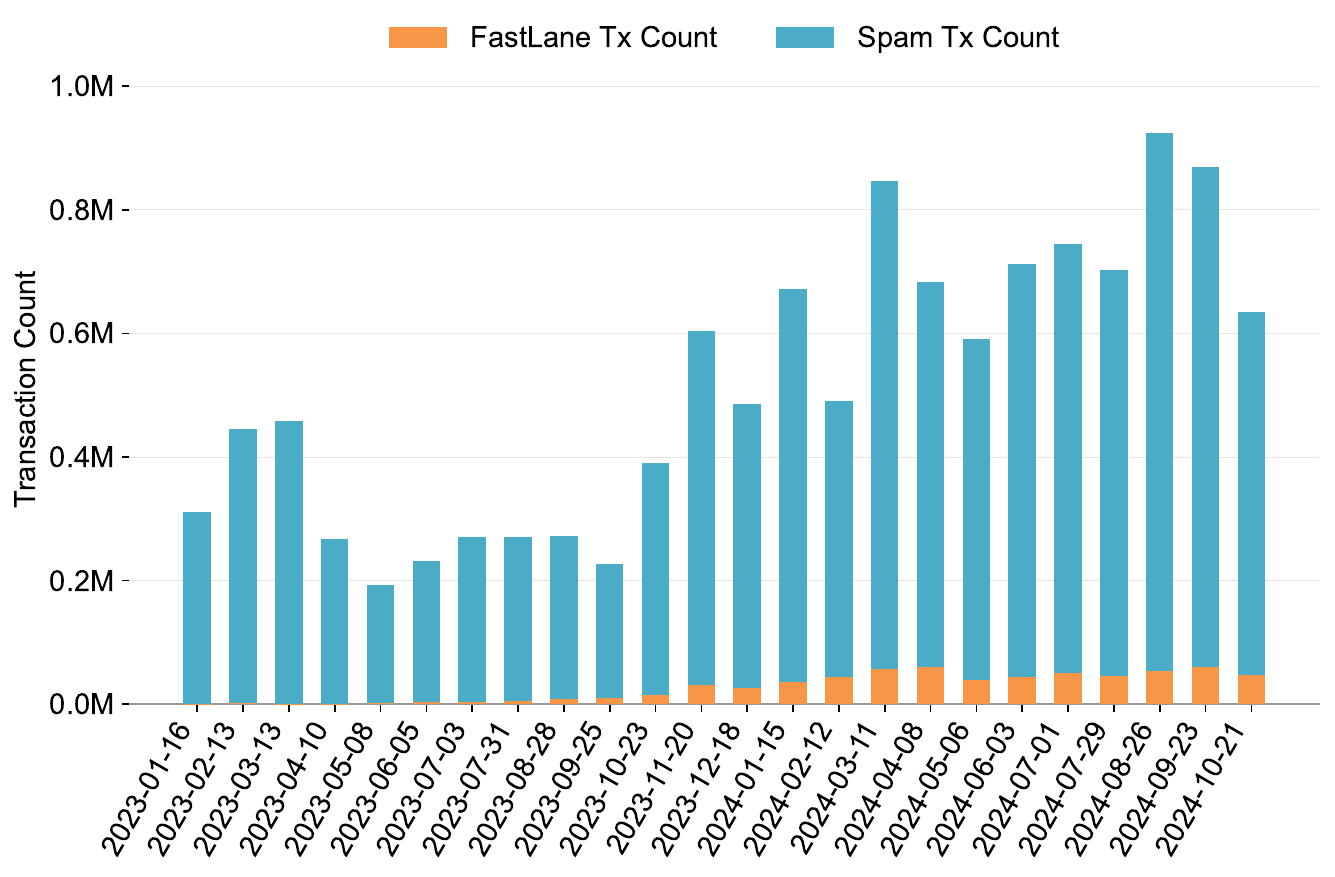}
    \caption{AA Transactions Count Over Time}
    \label{fig:AATxCntOvrTm}
\end{figure}

\begin{figure}[h]
    \centering
    \includegraphics[width=\columnwidth]{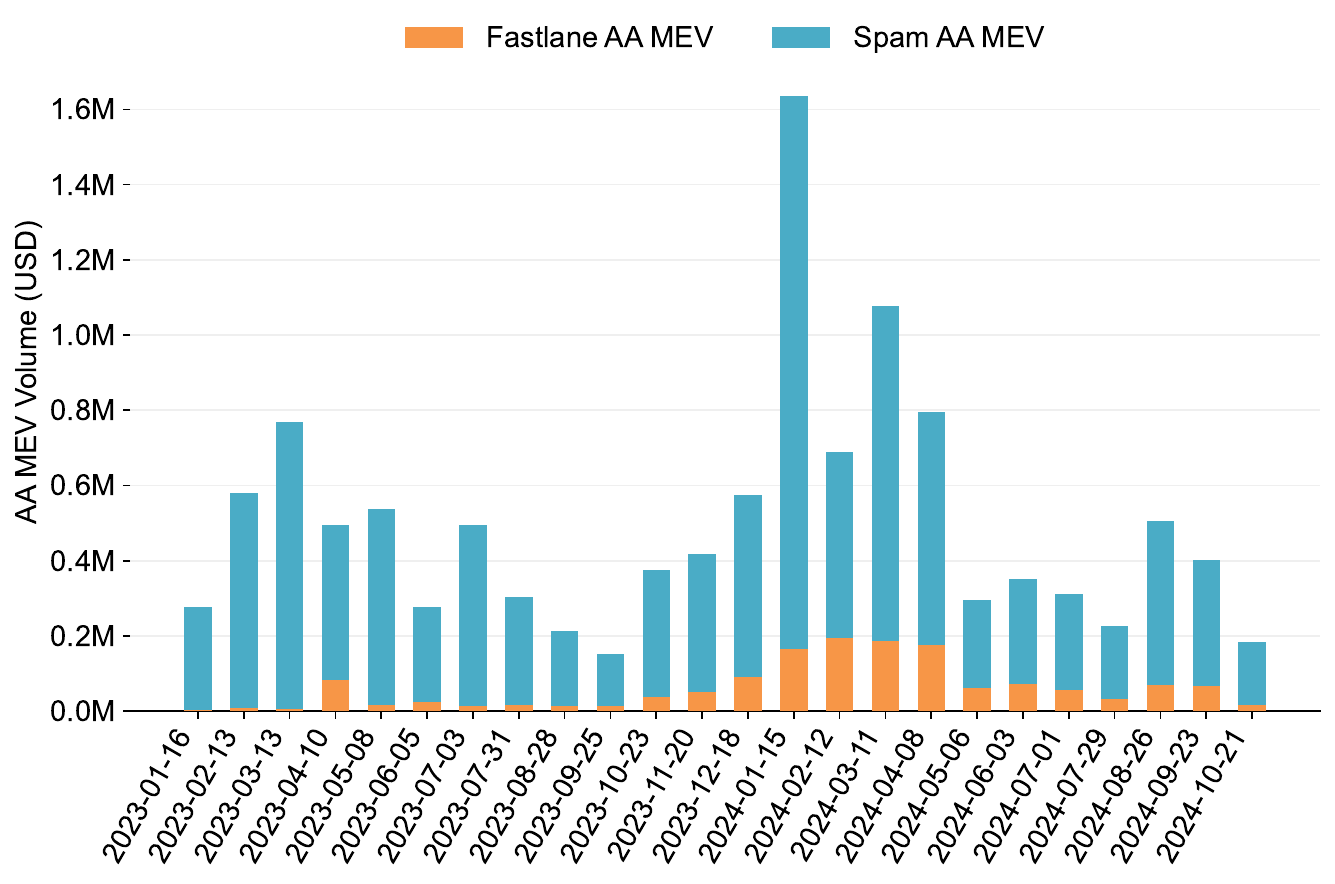}
    \caption{AA MEV Volume in USD}
    \label{fig:AAMEVVlmOrTm}
\end{figure}

Figure \ref{fig:AATxCntOvrTm} represents AA transaction activity over time, indicating the transaction count, while Figure \ref{fig:AAMEVVlmOrTm} illustrates the AA MEV volume, showing the total MEV amounts for each corresponding interval.

Over the examined period, the total MEV extracted from Polygon amounted to approximately 12M USD, derived exclusively from AA transactions. Our research has identified a notable increase in the adoption of structured MEV strategies, with Fast Lane now accounting for 13\% of total extracted AA MEV. The graphs further reveal rising FastLane shares in both MEV volume and transaction count, underscoring its efficiency in capturing value with reduced network strain. This shift highlights a growing recognition among searchers and validators of the benefits of optimized transaction execution.

\subsection{Trends}

\begin{figure}[h]
    \centering
    \includegraphics[width=\columnwidth]{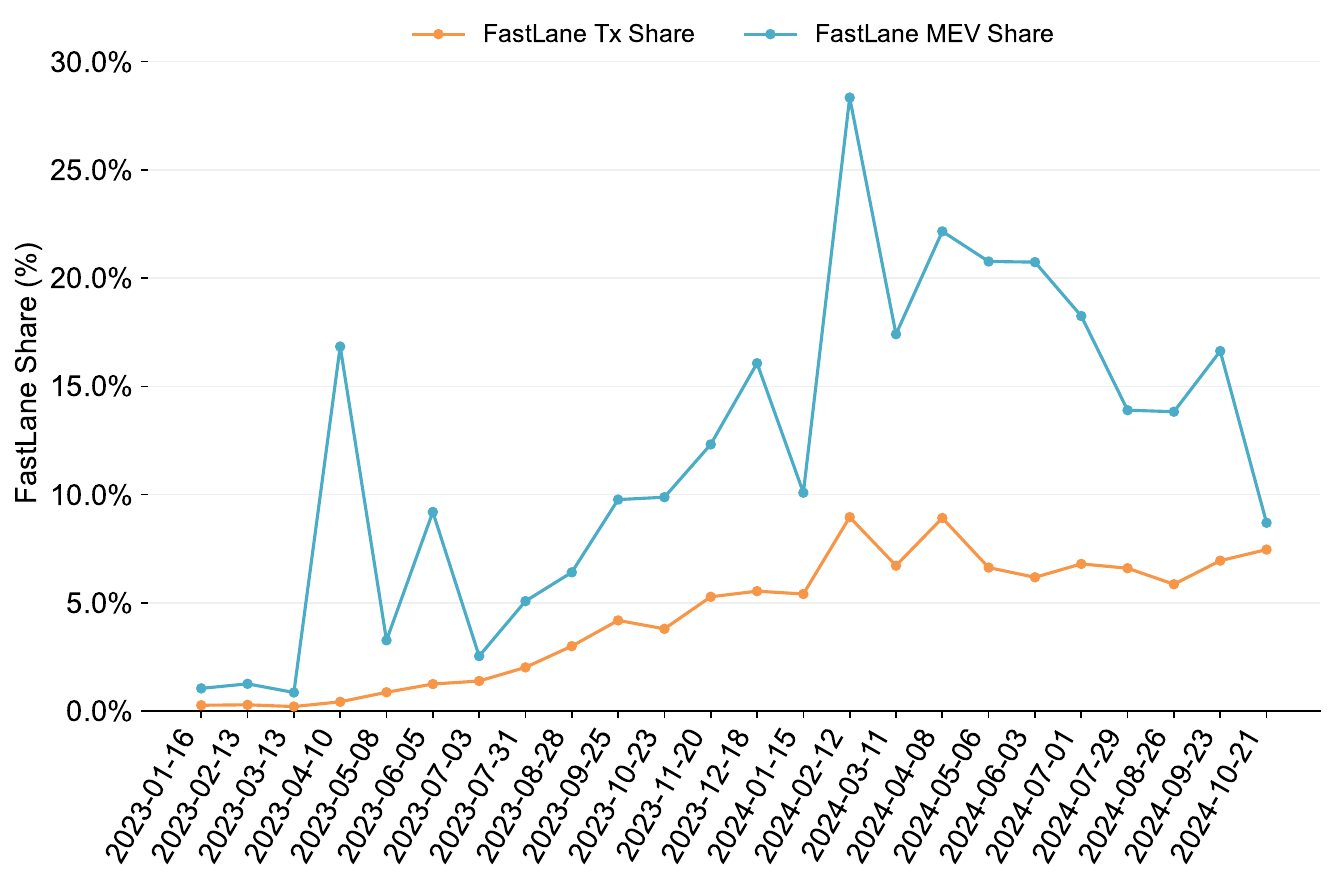}
    \caption{AA Fastlane Transactions and MEV Share Over Time}
    \label{fig:AAFlTxNMEVShrOvrTm}
\end{figure}

\begin{figure}[h]
    \centering
    \includegraphics[width=\columnwidth]{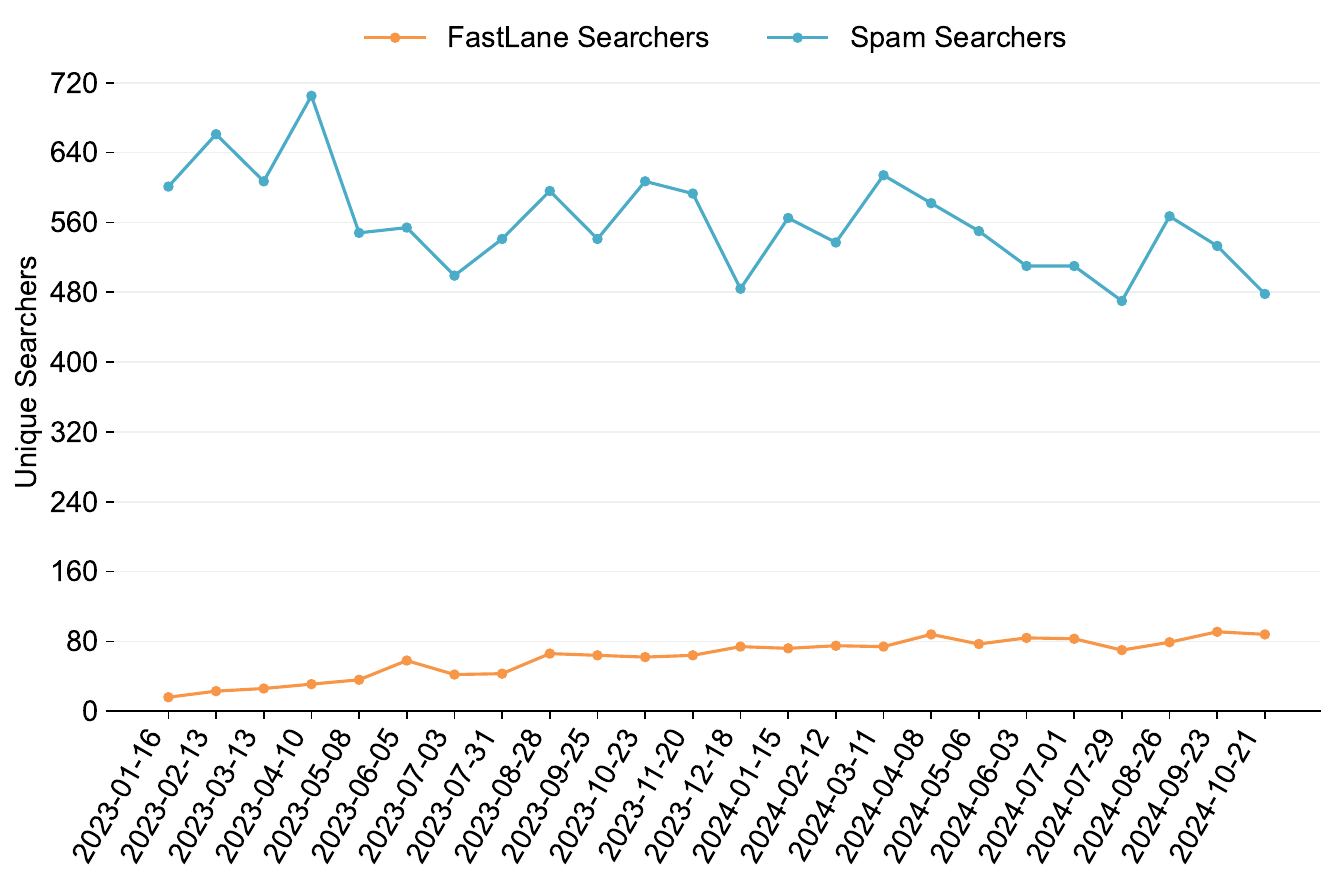}
    \caption{AA Unique Searchers Over Time}
    \label{fig:AAUqSrchrsOrTm}
\end{figure}

Figure \ref{fig:AAFlTxNMEVShrOvrTm} illustrates the share of FastLane transactions and MEV, revealing that while FastLane transactions constitute a small fraction of total transactions, they capture a significantly larger proportion of MEV. This suggests that FastLane transactions are more efficient than spam methods, achieving higher MEV extraction with fewer transactions. Complementing this, Figure \ref{fig:AAUqSrchrsOrTm} highlights the evolving landscape of unique searchers involved in FastLane and spam-based methods, showing a trend toward consolidation among spam searchers, which may indicate a more centralized and less competitive environment. In contrast, the increasing number of unique searchers adopting FastLane transactions suggests growing interest and participation in this approach, driven by its superior efficiency and profitability compared to traditional spam techniques.

\subsection{AA MEV Share}

\begin{figure}[h]
    \centering
    \includegraphics[width=\columnwidth]{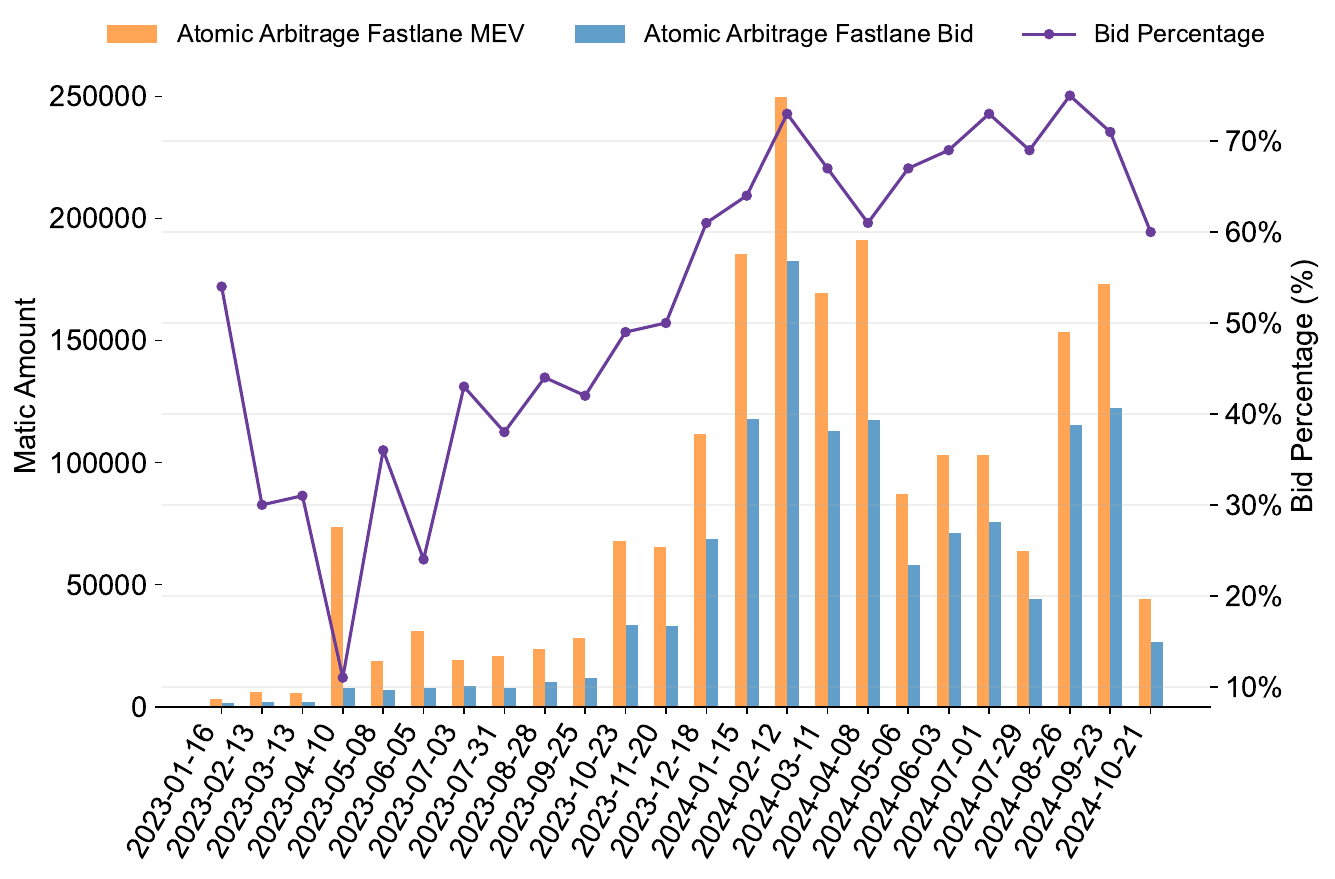}
    \caption{AA Transactions and Bids Share in FastLane Over Time}
    \label{fig:AAFlBidMEVBidtoMEVRtoOrTm}
\end{figure}

\begin{figure}[h]
    \centering
    \includegraphics[width=\columnwidth]{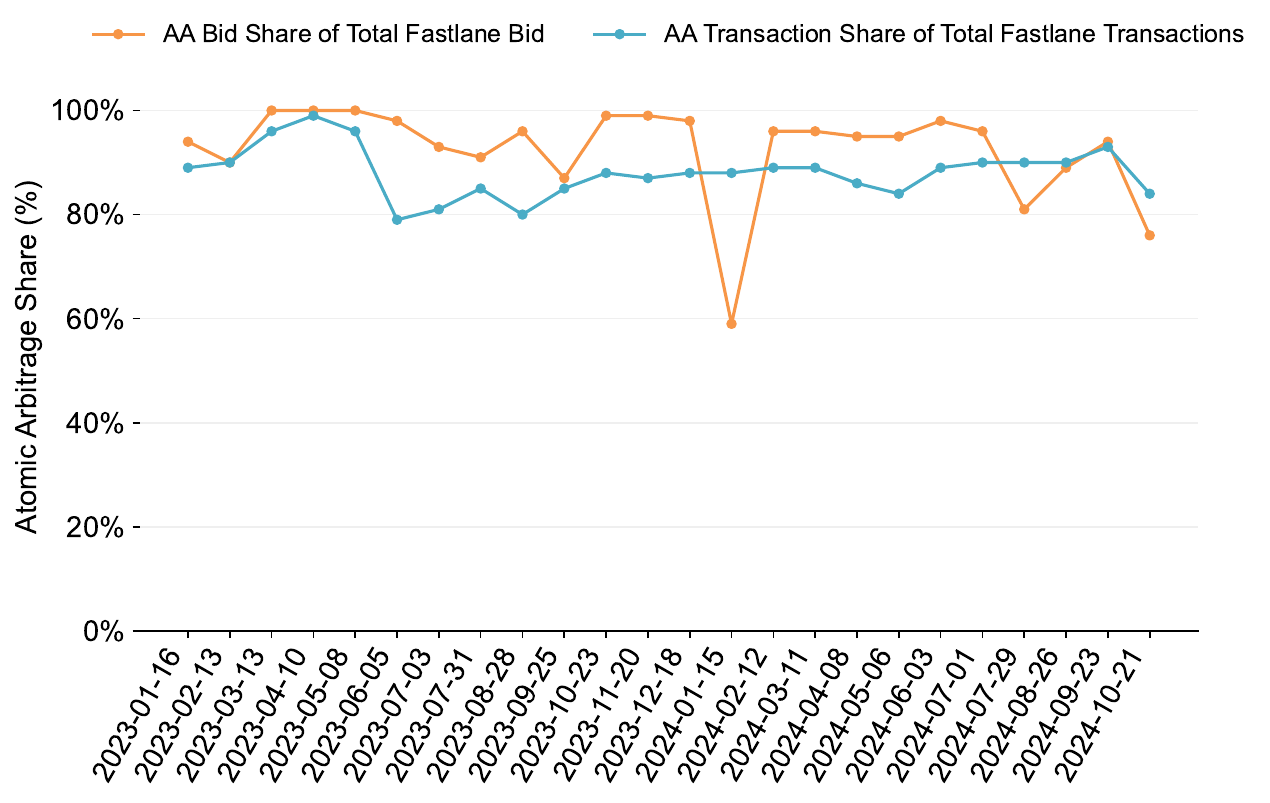}
    \caption{AA Fastlane Bid, MEV and Bid-to-MEV Ratio Over Time }
    \label{fig:AATxBidsShrFLOrTm}
\end{figure}

Figures \ref{fig:AAFlBidMEVBidtoMEVRtoOrTm} and \ref{fig:AATxBidsShrFLOrTm} illustrate trends in FastLane activity, covering bids, MEV volume, the Bid-to-MEV ratio, and the share of AA transactions and bids within total FastLane activity. Together, they reveal shifts in distribution, participation, and efficiency.

AA consistently accounts for approximately 90\% of backrunning-related MEV within the Polygon ecosystem. This conclusion is derived from the proportional relationship between AA bids and total FastLane bids, where AA strategies dominate transaction volume. The remaining 10\% of back running MEV is attributed to alternative strategies, including liquidations, NFT arbitrage, and protocol-specific exploits. Based on data from our research, the total estimated volume of back running MEV over the study period amounts to 13.5 million USD.
Notably, a distinct category of MEV,  top-of-block transactions (transactions prioritized in the initial positions of a block),  remains outside the scope of this analysis. These transactions represent a critical yet underexplored component of MEV dynamics and will be the focus of future research to comprehensively map the structure and volume of MEV across Polygon.

\section{Conclusion}\label{sec:Conclusion}

This paper presents a comprehensive study of MEV on the Polygon blockchain, focusing on AA transactions over a 22-month period. By categorizing AA transactions into Spam and FastLane based strategies, the paper highlight their distinct operational impacts. Spam transactions dominate in volume but significantly contribute to network congestion, while FastLane transactions, though fewer in number, extract a disproportionately large share of MEV through competitive bidding mechanisms.

The analysis reveals that MEV extraction is a substantial factor within the Polygon ecosystem, with approximately 1\% of the Total Value Locked (TVL) being captured as MEV. Validators emerge as the primary beneficiaries, receiving over 75\% of extracted MEV through direct bids and gas fees. The searchers, who execute the strategies, form the second group of beneficiaries, leveraging advanced techniques to compete for MEV opportunities. Further identifying an increasing shift towards structured and efficient MEV strategies, with FastLane transactions now accounting for 13\% of the total extracted AA MEV. This trend underscores the growing preference for optimized execution methods that reduce network strain while maximizing profitability. \added{These findings also raise concerns about fairness and decentralization, as MEV profits are concentrated among validators and sophisticated searchers. Ensuring equitable access to MEV opportunities and minimizing centralization risks is vital for the long-term health of Polygon’s DeFi ecosystem.} Moreover, the rise of intent-based transaction execution mechanisms suggests that the MEV landscape is evolving beyond extraction toward prevention.

It is anticipated that future developments in MEV protection will focus on minimizing inefficiencies at the transaction originator level. Emerging solutions such as CoW Swap, 1inch Fusion, and FastLane Atlas already aim to prevent MEV by restructuring transactions before they reach the mempool. This transformation could lead to a paradigm shift where MEV searchers evolve into solvers, actively optimizing transaction execution rather than exploiting inefficiencies. In this model, User-Opportunity Transactions (UserOpp TX) and Solver Operations (Solver Ops) will be bundled within a single transaction, preventing MEV opportunities from arising in the first place. Future work should expand the scope beyond AA transactions to include other MEV categories, such as liquidations and sandwich attacks, to develop a more comprehensive understanding of MEV dynamics. Additionally, examining MEV across multiple blockchain ecosystems with varying transaction ordering mechanisms will provide critical insights for designing fairer, more efficient, and decentralized financial infrastructures.

This study lays the groundwork for addressing MEV-related inefficiencies and shaping the next generation of blockchain design, where value extraction is balanced by mechanisms that enhance fairness, efficiency, and user protection.

\bibliographystyle{IEEEtran}
\bibliography{bibliography}

\end{document}